\documentclass[preprintnumbers,amsmath,11pt,amssymb,floatfix,showpacs,superscriptaddress,nofootinbib]{revtex4}
\usepackage{graphicx}
\usepackage{epsfig}
\usepackage{bm}
\usepackage{amsfonts}

\begin{document}

\title{\Large A study on the dependence of the dimensionless Brans-Dicke parameter on the scalar field and their time dependence}

\author{Sudipto Roy}
\email{roy.sudipto1@gmail.com}
\affiliation{Department of Physics,
St. Xavier's College,  Kolkata-700 016, India.}

\author{Surajit Chattopadhyay}
\email{surajit_2008@yahoo.co.in, surajcha@iucaa.ernet.in}
\affiliation{ Pailan College of Management and Technology, Bengal
Pailan Park, Kolkata-700 104, India.}

\author{Antonio Pasqua}
\email{toto.pasqua@gmail.com} \affiliation{Department of Physics,
University of Trieste, Via Valerio, 2 34127 Trieste, Italy.}

%\date{\today}

\begin{abstract}
\begin{center}
 \textbf{\textbf{Abstract}}
\end{center}
The present paper reports a study on the dimensionless parameter $\omega$  in Brans-Dicke theory. Based on a particular choice of scale factor $a$,  we have investigated the signature flip of the deceleration parameter $q$ to see whether the transition from decelerated to accelerated expansion of the universe is achievable under this choice of scale factor. Restrictions on the parameters obtained for this choice of scale factor have been subsequently used for
discussing the Brans-Dicke parameter for two choices of scalar fields $\phi$. Moreover, analytical solutions for the Brans-Dicke parameter without any assumption about the scalar field have been obtained from the modified field equations through the choice of scale factor under consideration. Viable models have been obtained by comparing the results with observations.

\end{abstract}


\maketitle

\newpage
\section{Introduction}
Riess et al. \cite{obs1} in the High-redshift
Supernova Search Team and Perlmutter et al. \cite{obs2} in the
Supernova Cosmology Project Team have independently reported that
the present universe is expanding with acceleration. Cosmological observations on expansion history of the universe can be interpreted as evidence
either for existence of some exotic matter components or for modification of the gravitational
theory. In the first route of interpretation one can take a mysterious cosmic fluid with sufficiently
large and negative pressure, dubbed dark energy. In the second route, however, one
attributes the accelerating expansion to a modification of general relativity \cite{BD9,BD10}. The
representative models belonging to the second class are known as
``modified gravity" models  \cite{mod1,mod2,mod3,mod4} which
include $f\left(R \right)$ gravity (with $R$ representing the Ricci scalar curvature) \cite{fr}, $f\left(T \right)$ gravity \cite{mioft1,mioft2} (where $T$ represents the torsion scalar), $f\left(G \right)$ gravity \cite{miog} (where $G = R^2 - 4R_{\mu \nu}R^{\mu \nu}+ R_{\mu \nu \lambda \sigma}R^{\mu \nu \lambda \sigma}$, with $R$ representing the Ricci scalar curvature, $R_{\mu \nu}$ representing the Ricci curvature tensor and $R_{\mu \nu \lambda \sigma}$ representing the Riemann curvature tensor), $f\left(R,T \right)$ gravity \cite{myr1,myr2,miofrt,miofrt2},  scalar-tensor theories
\cite{scalartensor} and braneworld models \cite{brane}.

Brans-Dicke (BD) theory is a special case of scalar-tensor theories, which is originally motivated by the search for a theory containing Mach's principle. The Brans-Dicke cosmology has been well studied considering different models. Sheykhi et al. \cite{bdcutoff1} considered the  power-law entropy-corrected version of BD theory defined by a scalar field $\phi$ and a coupling function $\omega$. As simplest and best-studied generalizations of General Relativity, we have the Holographic DE (HDE) and the New Agegraphic DE (NADE) models in the framework of BD cosmology. Sheykhi et al. \cite{bdcutoff2} considered the HDE model in BD theory to think about the BD scalar field as a possible candidate for producing cosmic acceleration without invoking auxiliary fields or exotic matter  considering the logarithmic correction to the entropy. Jamil et al. \cite{bdcutoff4} studied the cosmic evolution in Brans-Dicke chameleon cosmology. Pasqua $\&$ Khomenko \cite{bdcutoff5} studied the interacting logarithmic entropy-corrected HDE model in BD cosmology with IR cut-ff given by the average radius od the Ricci scalar curvature, i.e. $L= R^{-1/2}$. Pasqua $\&$ Chattopadhyay \cite{mionade} studied the main cosmological properties of the New Agegraphic DE (NADE) model in chameleon BD cosmology considering different expressions of the scale factor $a(t)$, in particular the emergent, the intermediate and the logamediate scale factors. Pasqua et al. \cite{miobdkhu} recently studied the main cosmological properties of the power law and logarithmic entropy corrected Ricci DE model in the framework of Brans-Dicke chameleon cosmology.  Different dark energy candidates have been considered in the framework of BD theory by \cite{setare1,setare2}.

In the present work, we are going to consider matter system to be a pressureless perfect fluid (dust) with energy density $\rho$ following \cite{NB,BD9}. Accelerated expansion is possible in BD theory in a matter-dominated universe as shown in \cite{BD10,anjan,sharif1}. Moreover, we are considering the possibility that the Newton's gravitational constant can vary with time. The Newton's gravitational constant $G$ has the role of a coupling between matter and geometry in the Einstein field equations. In an evolving universe, it becomes natural to consider the gravitational constant $G$ not anymore as "constant" but as a function of the time $t$.
Many suggestions based on different arguments  in which $G$ varies with time have been recently proposed.
Dirac \cite{new1,new1-1,new1-2,new1-3} was the first one to propose the idea of a time variable
$G$ on some physical grounds. He showed that $G$ has the following time dependance: $G \left( t \right) \propto t^{-1}$. However, this model lead to some difficulties.  Abdel Rahaman \cite{new2} and Mass \cite{new3} have demonstrated that $G$ can be described as an increasing function of the time $t$. Many other extensions of Einstein's theory with a time dependent gravitational constant  $G$ have also been recently considered in order to obtain a possible unification between the theory of gravitation and the theory of  elementary particle physics or to incorporate Mach's principle in General Relativity \cite{new4,new5,new6}. Canuto $\&$ Narlikar \cite{new7} showed that cosmology with time variable $G$ results to be consistent with  cosmological observations presently available. Some constraints on  the value of $\dot{G}/G$ (with $\dot{G}$ being the time derivative of $G$) can be obtained from different sources. According to  large numbers hypothesis  discovered by Weyl, Eddington and Dirac, $\dot{G}/G$ goes as the Hubble rate $H$ \cite{new8}. Observations of the Hulse-Taylor binary pulsar B1913+16 indicate the estimate $0 < \dot{G}/G < \left(2 \pm 4\right)\times  10^{-12}yr^{-1} $ \cite{new9}. Moreover, helioseismological data suggest $0 < \dot{G}/G <1.6\times 10^{-12}yr^{-1} $ \cite{new10,new11}. Recent works on the time variable gravitational constant  $G$  include \cite{new12,new13,new14,new15,new16,new17,new18}.

The present paper is organized as follows. In Section II, we have considered a special form of scale factor $a$ and studied its consequences on the accelerating universe. In Section III, we have formulated the BD parameter $\omega$ as a function of the scalar field $\phi$. In this connection, we have considered three different models. Finally, in Section IV, we have presented the Conclusions of this work.

\section{\textbf{A scale factor causing signature flip of deceleration parameter}}
The deceleration parameter $q$ is defined as:
\begin{eqnarray}\label{1}
q=-\frac{a \ddot{a}}{\dot{a}^2},
\end{eqnarray}
where $a\left(t\right)$ represents the scale factor, which gives information about the expansion of the universe, and an overdot stands for a time derivative. Other expression of $q$, however, can be found in literature. The accelerated expansion of the present universe is well-documented in literature. Any change of state of motion of the universe from decelerated to accelerated phase should cause the deceleration parameter $q$ to change its sign from positive to negative. The functional form of $q$ in Eq. (\ref{1}) clearly indicates that it can change sign (say from $+ve$ to $-ve$), as a function of time, only if $\ddot{a}$ changes sign in the opposite manner (say from $-ve$ to $+ve$). Let us choose a simple functional form for the scale factor $a \left(t\right)$ such that its double derivative $\ddot{a}$ shows a signature flip at a certain instant of time $t$. For this purpose, let us take the following functional form of the scale factor
\begin{eqnarray}\label{2}
a\left(t\right)=At^n\exp[bt],
\end{eqnarray}
where $A,~n$ and $b$ are positive constants considering $a \left(t\right)$  and $\dot{a}(t)$  to be positive quantities. Using the above functional form of the scale factor, the deceleration parameter comes out to be:
\begin{eqnarray}\label{3}
q\left(t\right)=-1+\frac{n}{(n+bt)^2}.
\end{eqnarray}
Consequently, $-1<q\leq\frac{1}{n}$. It will be shown later that $n<1$. The instant of time (say $t_1$) at which the above expression of $q$ becomes zero is given by:
\begin{eqnarray}\label{4}
t_1=\frac{\sqrt{n}-n}{b}~~\Rightarrow~~b=\frac{\sqrt{n}-n}{t_1}.
\end{eqnarray}
In order $\dot{a}$ is positive, we must have $b>0$, which implies $\sqrt{n}>n$, leading to $n<1$.
Thus, to have a signature flip of $q\left(t\right)$, as defined by Eq. (\ref{3}), we must have $n<1$ in the expression of $a \left(t\right)$ we have chosen (which is given in Eq. (\ref{2})). Eq. (\ref{4}) precisely gives us the instant at which $\ddot{a}$ changes its sign from negative to positive.

We here assume that at the point of time $t_1$, when the universe transits from deceleration to acceleration, the value of $a \left(t\right)$ is $a_1$, i.e. at $t=t_1$, we have $a \left(t\right)=a_1$ and $q\left(t\right)=0$. Hence, Eq. (\ref{2}) can be rewritten as:
\begin{eqnarray}\label{5}
a \left(t\right)=a_1\left(\frac{t}{t_1}\right)^n\exp[b(t-t_1)].
\end{eqnarray}
Substituting the expression of $b$ given in Eq. (\ref{4}) in Eqs. (\ref{3}) and (\ref{5}), the expressions for the scale factor $a\left( t \right)$ and the deceleration parameter $q$ become, respectively:
\begin{eqnarray}
a \left(t\right)&=&a_1\left(\frac{t}{t_1}\right)^n\exp\left[(\sqrt{n}-n)\left(\frac{t}{t_1}-1\right)\right], \label{6}\\
q\left(t\right)&=&-1+n\left[n+(\sqrt{n}-n)\frac{t}{t_1}\right]^{-2}. \label{7}
\end{eqnarray}
In Eq. (\ref{7}), it is easy to show that, for $t>t_1$, we have $q<0$ while, for $t<t_1$, we have $q>0$. Using Eqs. (\ref{6}) and (\ref{7}), we can express the scale factor as a function of deceleration parameter $q$ as follow:
\begin{eqnarray}\label{8}
a\left(q\right)=a_1\left[\frac{1}{\sqrt{n}-1}\left\{\left(\frac{q+1}{n}\right)^{-\frac{1}{2}}-n\right\}\right]^n\exp\left[\left(\frac{q+1}{n}\right)^{-\frac{1}{2}}-\sqrt{n}\right].
\end{eqnarray}
The cosmological redshift $z$ is related to the scale factor $a\left( t \right)$ by the following relation:
\begin{eqnarray}\label{9}
z=a^{-1}-1.
\end{eqnarray}
Using Eq. (\ref{8}) in Eq. (\ref{9}), we get:
\begin{eqnarray}\label{10}
z=-1+\frac{1}{a_1}\left[\frac{1}{\sqrt{n}-1}\left\{\left(\frac{q+1}{n}\right)^{-\frac{1}{2}}-n\right\}\right]^{-n}\exp\left[\sqrt{n}-\left(\frac{q+1}{n}\right)^{-\frac{1}{2}}\right].
\end{eqnarray}
It is evident from Eq. (\ref{10}), that for $q=0$, we have $z=-1+\frac{1}{a_1}$. Thus, we see that the deceleration parameter $q$ crosses zero value in favour of a negative one at $z=-1+\frac{1}{a_1}$.

 Eqs. (\ref{6}) and (\ref{7}) make it clear that the evolution of $a \left(t\right)$ and $q \
 \left(t\right)$ can be described with respect to a relative measure of time $t_r=\frac{t}{t_1}$. We assume that $q_0$ denotes the deceleration parameter at the present epoch i.e. $a=1$. Taking $n=0.89$ in  Eqs. (\ref{6}) and (\ref{7}) we find that at $t_r=2.57$  we have $a=1.01$  and $q=-0.16$. It means that for a parametric value of $n=0.89$, $q=0$ at $a=\frac{2}{5}$ (i.e. $z=1.5$) and hence $q_0\thickapprox-0.16$ . These values are quite consistent with those obtained in the study of \cite{NB}. Considering a different case where $q$ crosses zero at $z=1$ (i.e. $a=a_1=\frac{1}{2}$),  we have $q_0\thickapprox-0.12$ at $t_r=2.07$.  Assuming the signature flip in $q$ to have taken place at $z=0.5$  (i.e.$a=a_1=\frac{2}{3}$), we get $q_0\thickapprox-0.06$ at $t_r=1.53$. These values for $z=0.5$ and $1$, have been calculated keeping the parameter $n$ fixed at $0.89$.

 We now define a new time parameter $T_r=\frac{t}{t_0}$, where $t_0$ is the present cosmic time where $a=a_0=1$. Hence we have $t_r=\left(\frac{t_0}{t_1}\right)T_r$, where $t_1$ is the time when $q=0$. To obtain the value of $\frac{t_0}{t_1}$  from Eq. (\ref{7}) we have to substitute here the value of $q$ at $a=1$. In order to do this, we need an expression of the deceleration parameter $q$ explicitly in terms of the scale factor $a$. Using Eq. (\ref{8}), we have determined such a relation numerically for $n=0.89$. Over the range of values of $q$ going from $0.1$ to $-0.2$, we have numerically found that:
 \begin{eqnarray}
q\left(a\right)=C+D\exp\left[-\frac{a}{f\left(a_1\right)}\right], \label{11}
 \end{eqnarray}
 where
\begin{eqnarray}\label{12}
 f\left(a_1\right)=\xi+\eta a_1,
\end{eqnarray}
with $C=-0.606,~~D=0.737,~~\xi=2.222\times10^{-6}~\textrm{and}~\eta=5.097$. This is a kind of a semi-empirical expression of $q(a)$ that depends upon the value of $a_1$, which is the value of the scale factor $a$ at $q=0$. In Eq. (\ref{12}) the value of $\xi$ is six orders of magnitude smaller than that of $m$. Hence, $\xi$ can be safely neglected when $a_1$ is not too small. Eq. (\ref{12}) clearly shows that:
\begin{eqnarray}\label{13}
q_0=C+D\exp\left[-\frac{1}{\xi+\eta a_1}\right].
\end{eqnarray}
Putting $a_1=\frac{2}{5},~\frac{1}{2}~\textrm{and}~\frac{2}{3}$ in Eq. (\ref{13}) we get respectively $q_0=-0.15,~-0.11~\textrm{and}~-0.06$. These values are quite close to the values calculated earlier. Using Eq. (\ref{11}) in Eq. (\ref{7}) we get:
\begin{eqnarray}\label{14}
C+D\exp\left[-\frac{a}{f(a_1)}\right]=-1+n\left\{n+(\sqrt{n}-n)\frac{t}{t_1}\right\}^{-2}.
\end{eqnarray}
For the present epoch, we have $t=t_0$ and $a=1$. Using the definition $T_r=\frac{t}{t_0}$ in Eq. (\ref{14}) we get for the present epoch:
\begin{eqnarray}\label{16}
t_r=T_r\frac{1}{\sqrt{n}-n}\left[\left(\frac{1+C+D\exp\left[-\frac{1}{f(a_1)}\right]}{n}\right)^{-\frac{1}{2}}-n\right].
\end{eqnarray}
Using Eq. (\ref{16}) in Eqs. (\ref{6}) and (\ref{7}), we get:
\begin{eqnarray}
a \left(t\right) &=& a_1\left[T_rF(a_1)\right]^n\exp\left[(\sqrt{n}-n)(T_rF(a_1)-1)\right], \label{17}\\
q\left(t\right) &=& -1+n\left\{n+(\sqrt{n}-n)F(a_1)T_r\right\}^{-2}, \label{18}
\end{eqnarray}
where:
\begin{eqnarray}\label{19}
F\left(a_1\right)=\frac{1}{\sqrt{n}-n}\left[\left(\frac{1+C+D\exp\left[-\frac{1}{f(a_1)}\right]}{n}\right)^{-\frac{1}{2}}-n\right].
\end{eqnarray}
Eqs. (\ref{17}) and (\ref{18}) allows us to calculate the scale factor $a$ and the deceleration parameter $a$ respectively as a function of a time parameter $T_r=\frac{t}{t_0}$, measured with respect to the present epoch $t_0$.
From Eqs. (\ref{2}) and (\ref{4}), we have:
\begin{eqnarray}\label{20}
H=\frac{\dot{a}}{a}=\frac{\sqrt{n}-n}{t_1}+\frac{n}{t}.
\end{eqnarray}
Using Eqs. (\ref{3}) and (\ref{20}), we have:
\begin{eqnarray}\label{22}
\frac{H}{H_0}=\frac{1+n\left\{\left(\frac{q+1}{n}\right)^{-\frac{1}{2}}-n\right\}^{-1}}{1+n\left\{\left(\frac{q_{0}+1}{n}\right)^{-\frac{1}{2}}-n\right\}^{-1}}.
\end{eqnarray}
Using Eq. (\ref{11}) in Eq. (\ref{22}) we get:
\begin{eqnarray}\label{23}
\frac{H}{H_0}=\frac{1+n\left\{\left(\frac{C+D\exp[-a/f(a_1)]+1}{n}\right)^{-\frac{1}{2}}-n\right\}^{-1}}{1+n\left\{\left(\frac{C+D\exp[-1/f(a_1)]+1}{n}\right)^{-\frac{1}{2}}-n\right\}^{-1}}.
\end{eqnarray}
Putting $H=H_0$ at $t=t_0$ in Eq. (\ref{20}) we have:
\begin{eqnarray}\label{24}
H=H_0-\frac{n}{t_0}+\frac{n}{t}.
\end{eqnarray}
Using Eq. (\ref{24}) in (\ref{2}) we get for $t=t_0$:
\begin{eqnarray}\label{25}
a \left(t\right)=\left(\frac{t}{t_0}\right)^n\exp\left[(H_0t_0-n)\left(\frac{t}{t_0}-1\right)\right].
\end{eqnarray}
Since $H_0=b+\frac{n}{t_0}$ we get from Eq. (\ref{3}) that:
\begin{eqnarray}\label{26}
q\left(t\right)=-1+n\left[n+(H_0t_0-n)\frac{t}{t_0}\right]^{-2}.
\end{eqnarray}
It is evident from Eq. (\ref{26}) that in order to have $q<0$ at $t=t_0$ we require $n<H_0^2t_0^2$. Using the fact that $q=0$ at $t=1$ we have from Eq. (\ref{26}) that:
\begin{eqnarray}\label{27}
\frac{t_1}{t_0}=\frac{\sqrt{n}-n}{H_0t_0-n}.
\end{eqnarray}
From Eq. (\ref{27}) we get $H_0t_0\rightarrow\sqrt{n}$ as $t_1\rightarrow t_0$. Since $q=q_0$ at $t=t_0$, we can get from Eq. (\ref{26}) that:
\begin{eqnarray}\label{28}
n=(H_0t_0)^2(q_0+1).
\end{eqnarray}
Using Eqs. (\ref{25}) and (\ref{27}), we get the expression for $a_1$ as:
\begin{eqnarray}\label{29}
a_1=\left(\frac{\sqrt{n}-n}{H_0t_0-n}\right)^n\exp(\sqrt{n}-H_0t_0).
\end{eqnarray}
\begin{figure}
\centering
\includegraphics[height=8.2cm]{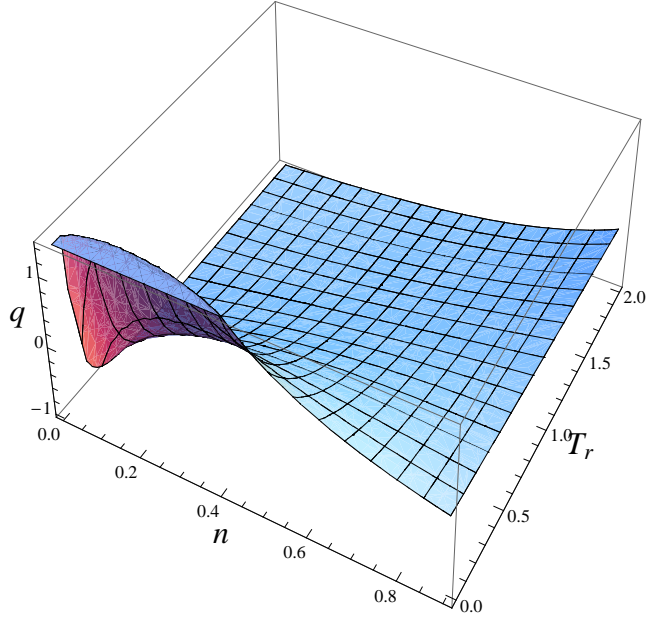}
\caption{Plot of $q$ against $T_r$ and $n$. Eqs. (\ref{11}), (\ref{12}), (\ref{25}) and (\ref{29}) have been used with $H_0t_0=0.95$ and $0<n<(H_0t_0)^2=0.9025$.}
\label{q}
\end{figure}
\begin{figure}
\centering
\includegraphics[height=8.2cm]{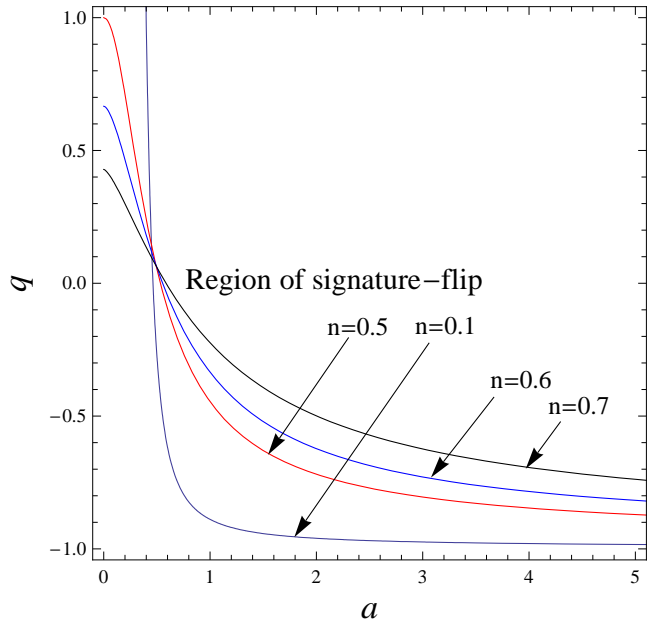}
\caption{Plot of the deceleration parameter $q$ against the scale factor $a$ using Eqs. (\ref{11}), (\ref{12}), (\ref{25}) and (\ref{29}) with $H_0t_0=0.95$ and $0<n<(H_0t_0)^2=0.9025$. We observe that as we increase the value of $n$, the signature-flip of $q$ occurs at higher values of $a$. Furthermore,  for larger values of $n$ we are getting smaller negative values of $q$ at present epoch $(a=1)$.}
\label{aq}
\end{figure}

In Fig. \ref{q} we have plotted the deceleration parameter $q$ with $0\leq t\leq4$ and $0<n<1$ using Eqs. (\ref{11}), (\ref{12}), (\ref{25}) and (\ref{29}). We have chosen $H_0t_0=0.95$. In the 3D-plot, a clear signature-flip of the deceleration parameter $q$ is observed for all values on $n$ under consideration. However, it may be noted that for smaller values of $n$ the signature flip is occurring at an earlier stage than that for the higher values of $n$.
In Fig. \ref{aq} we have parametrically plotted the deceleration parameter $q$ against the scale factor $a$ for the same set of values as in Fig. \ref{q}. It is evident from the plot that, for $a=1$, we have $q<0$. Thus, the present acceleration of the universe is achieved. Furthermore, this figure shows that the signature flip is happening roughly in the range $0.5<a<1$. This is consistent with the present accelerated universe. For some values of $n$ and for $a=1$ we have $q\approx -0.2$. This result is consistent with the study of Giostri et al. \cite{Giostri}, which states that combining BAO/CMB observations with SN Ia data processed with the MLCS2k2 light-curve fitter gives $q_0=-0.31_{-0.11}^{+0.11}$ at $68\%$ confidence level.
\\
\section{Formulation of BD parameter as a function of scalar field}
Within the framework of the Friedmann-Robertson-Walker (FRW) cosmology, the line element for a non-flat universe can be written, in polar coordinates, as follows:
\begin{eqnarray}
    ds^2=-dt^2+a^2\left(t\right)\left[\frac{dr^2}{1-kr^2} +r^2 \left(d\theta ^2 + \sin^2 \theta d\xi ^2\right) \right], \label{66}
\end{eqnarray}
where $t$ represents the cosmic time, $r$ is the radial component, $k$ is the curvature parameter and $\theta$ and $\xi$ are the two polar coordinates. \\
The action $S_{BD}$ of the BD cosmology is given by \cite{NB1}:
\begin{eqnarray}
S_{BD}=\int d^4x\sqrt{g}\left( -\varphi R+\frac{\omega}{\varphi} g^{\mu \nu}\partial_{\mu}\varphi \partial_{\nu}\varphi + L_m    \right). \label{77}
\end{eqnarray}
Defining $\varphi$ as:
\begin{eqnarray}
\varphi = \frac{\phi^2}{8\omega}, \label{88}
\end{eqnarray}
the action $S_{BD}$ given in Eq. (\ref{77}) can be rewritten in its canonical form:
\begin{eqnarray}
    S_{BD}=\int d^4x\sqrt{g}\left( -\frac{1}{8\omega}\phi^2R+\frac{1}{2}g^{\mu \nu}\partial_{\mu}\phi \partial_{\nu}\phi+L_m    \right), \label{99}
\end{eqnarray}
where $g$, $\omega$, $\phi$, $R$ and $L_m$ represent, respectively, the determinant of the tensor metric $g^{\mu \nu}$, the BD parameter, the BD scalar field, the Ricci scalar curvature and the Lagrangian of the matter. For flat FRW universe (which corresponds to a curvature parameter $k$ equal to zero), the field equations in the generalized BD theory are given by:
\begin{eqnarray}
3H^2&=&\frac{\rho}{\phi}+\frac{\omega\left(\phi\right)}{2}\left(\frac{\dot{\phi}}{\phi}\right)^2-3H\frac{\dot{\phi}}{\phi}, \label{31} \\
2\frac{\ddot{a}}{a}+H^2&=&-\frac{\omega\left(\phi\right)}{2}\left(\frac{\dot{\phi}}{\phi}\right)^2-2H\frac{\dot{\phi}}{\phi}-\frac{\ddot{\phi}}{\phi}, \label{32}
\end{eqnarray}
where  $\rho$ represents the energy density of the matter distribution and an overdot indicates a derivative with respect to the cosmic time $t$.
At this juncture, we should briefly discuss the issue of taking $\omega$ as a function of $\phi$ instead of a constant. This issue is elaborately discussed in the work of Das and Banerjee \cite{NB1} and has been further studied in \cite{NB}. This approach was earlier adopted in \cite{omegaphi1}, where it was shown that the Brans-Dicke scalar field interacting with dark matter can indeed generate an acceleration, where $\omega$ is not restricted to low values, but the
parameter $\omega$ had to be taken as a function of the scalar field $\phi$. In the presented work we have went along the line adopted and thoroughly explained in \cite{NB,NB11,omegaphi1}. The scalar field $\phi$ will subsequently be considered as a function of cosmic time $t$, and the work of \cite{sahoo} has treated $\omega$ as a function of $t$.
\\
In the following subsections we will consider three different cases in order to study their properties.
\subsection{Model-I}
In this subsection, we choose the scalar field $\phi$ as function of the scale factor $a$ expressed in the following form:
\begin{eqnarray}\label{33}
\phi\left(a\right)=\phi_1\exp[\alpha a].
\end{eqnarray}
Combining Eqs. (\ref{31}) and (\ref{32}) and using Eq. (\ref{33}), we get:
\begin{eqnarray}\label{34}
\left(\frac{2}{a}+\alpha\right)\ddot{a}+\left(\frac{4}{a^2}+\frac{5\alpha}{a}+\alpha^2\right)\dot{a}^2=\frac{\rho}{\phi},
\end{eqnarray}
Eq. (\ref{34}) can be rewritten as:
\begin{eqnarray}\label{35}
\alpha^2 a^2+Q(\alpha a)+R=0,
\end{eqnarray}
where  $Q=5-q$ (with $q$ being deceleration parameter) and $R=4-2q-\frac{\rho}{\phi H^2}$. The thermodynamic pressure of the cosmic fluid
is taken to be zero consistently with the present dust universe \cite{NB}. Hence, we assume
the conservation equation for matter leading to the relation:
\begin{eqnarray}\label{36}
\rho=\rho_0 a^{-3},
\end{eqnarray}
where $\rho_0$ is a constant indicating the present day value of $\rho$. Taking $\phi=\phi_0$ at $a=a_0=1$, we get from Eq. (\ref{33})
\begin{eqnarray}\label{37}
\phi(a)=\phi_{0} \exp[\alpha(a-1)].
\end{eqnarray}
Using Eqs. (\ref{28}), (\ref{36}) and (\ref{37}) in Eq. (\ref{35}), we get, for $a_0=1$:
\begin{eqnarray}\label{39}
\alpha^2+\left[6-\frac{n}{H_0^2t_0^2}\right]\alpha+\left[6-\frac{2n}{H_0^2t_0^2}-\frac{f_0t_0^2}{H_0^2t_0^2}\right]=0;~~~\textrm{where}~~f_0=\frac{\rho_0}{\phi_0}.
\end{eqnarray}
The CMB measurements \cite{obs1} put $1.05$ as upper limit of the value of $H_0t_0$ \cite{krs}. For this reason, we consider $H_0t_0=f_1<1$. It has been already established in the previous Section that, in order to have a signature flip, we require $n<H_0^2t_0^2=f_1^2<1$. Hence, we can have $f_2<1$ such that $n=f_2f_1^2$. We further define the parameter $f_3$ as follow:
\begin{eqnarray}\label{44}
f_3=f_0t_0^2.
\end{eqnarray}
Hence, Eq. (\ref{39}) takes the following form:
\begin{eqnarray}\label{45}
\alpha^2+(6-f_2)\alpha+\left(6-2f_2-\frac{f_3}{f_1^2}\right)=0,
\end{eqnarray}
which is quadratic in $\alpha$. The two roots of Eq. (\ref{45}) are:
\begin{eqnarray}\label{46}
\alpha_{\pm}=\frac{f_2-6\pm\sqrt{f_2^2-4f_2+4f_3/f_1^2+12}}{2}.
\end{eqnarray}
Let us denote the scalar fields corresponding to $\alpha_+$ and $\alpha_-$, respectively, as:
\begin{eqnarray}
\phi_{+}\left(a\right) &=& \phi_0\exp[\alpha_{+}(a-1)],  \label{47} \\
\phi_{-}\left(a\right)&=& \phi_0\exp[\alpha_{-}(a-1)]. \label{48}
\end{eqnarray}

\begin{figure}[ht] \begin{minipage}[b]{0.45\linewidth} \centering\includegraphics[width=\textwidth]{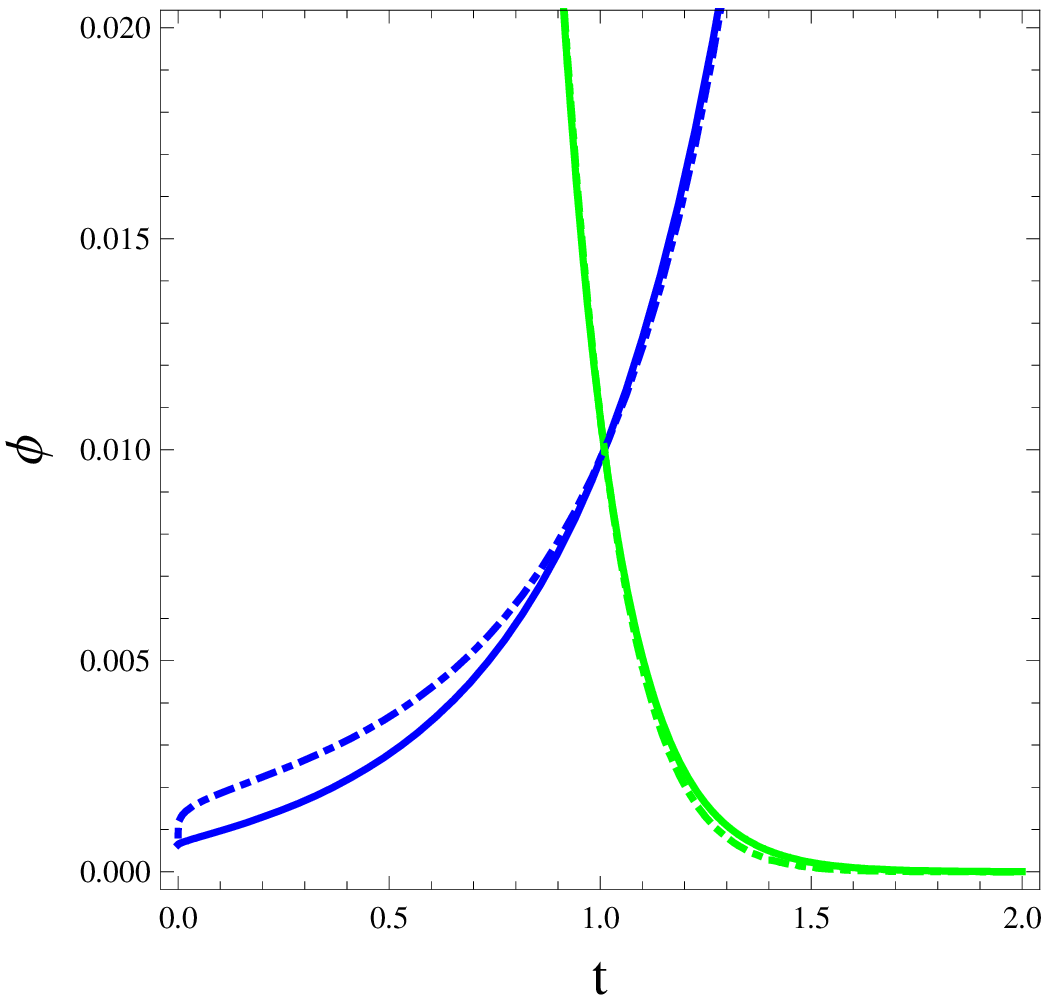} \caption{Plot of $\phi$ (see Eqs. (43) and (44)) against the cosmic time $t$ in Model I.} \label{phi} \end{minipage} \hspace{0.5cm} \begin{minipage}[b]{0.45\linewidth} \centering\includegraphics[width=\textwidth]{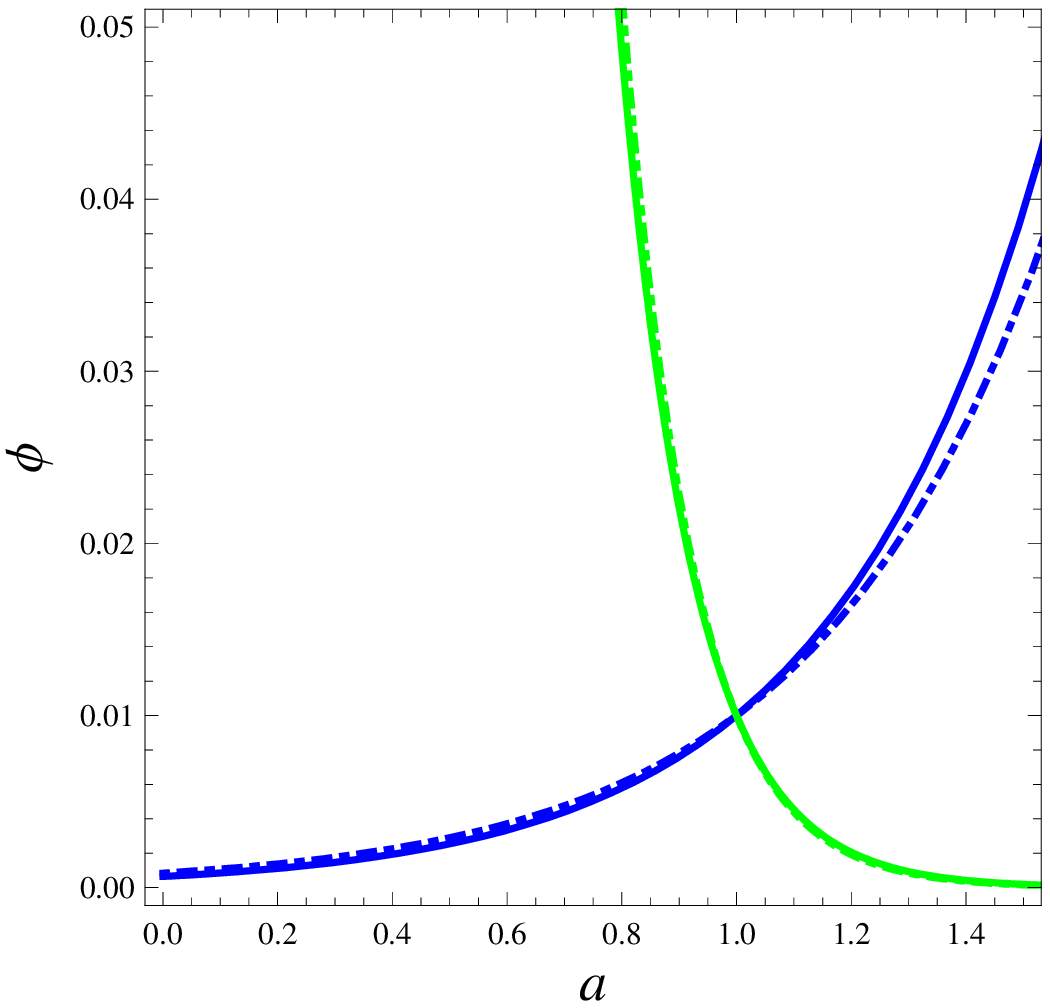} \caption{Plot of $\phi$ (see Eqs. (43) and (44)) against the scale factor $a$ in Model I.} \label{phia1} \end{minipage} \hspace{0.5cm} \begin{minipage}[b]{0.45\linewidth} \centering\includegraphics[width=\textwidth]{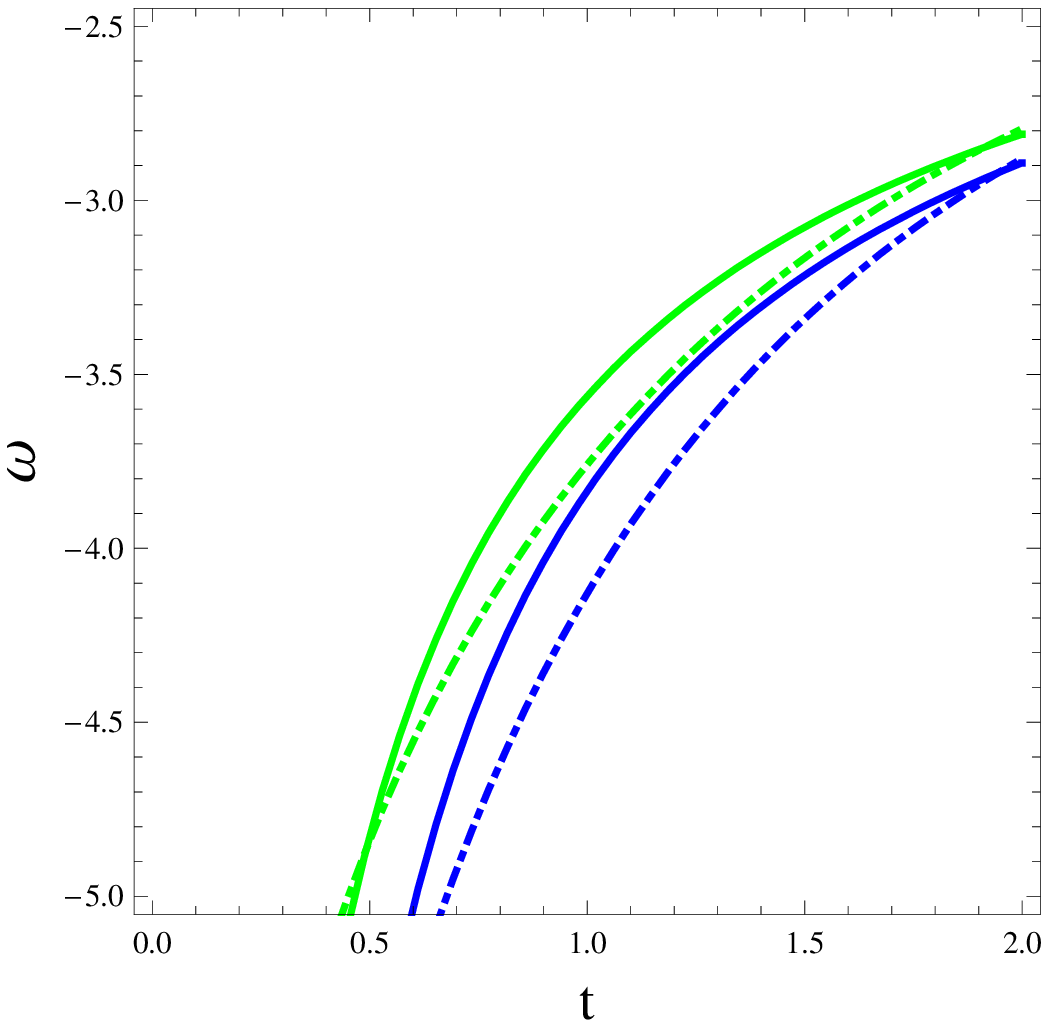}\caption{Plot of BD parameter $\omega$ (see Eq. (45)) against the time $t$ in Model I.} \label{omega} \end{minipage}\hspace{0.5cm}\begin{minipage}[b]{0.45\linewidth} \centering\includegraphics[width=\textwidth]{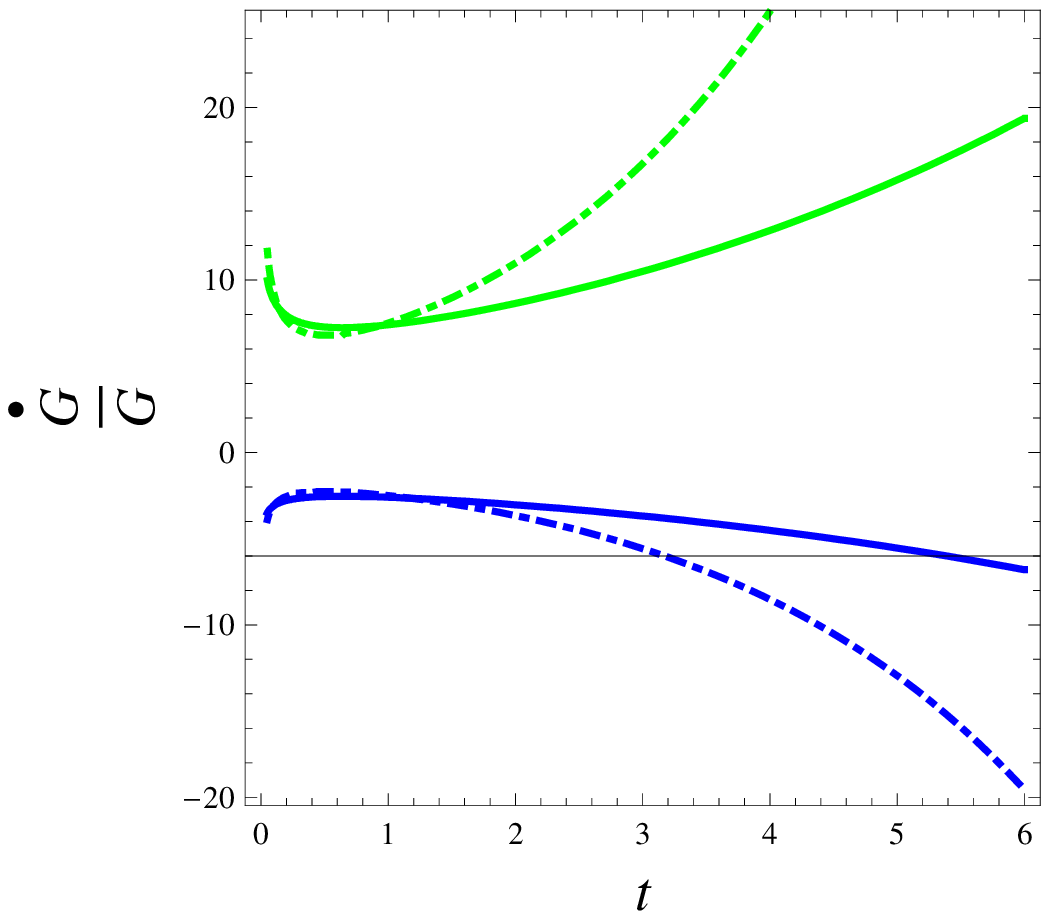}\caption{Plot of time derivative $\dot{G}$ of $G$ (see Eq. (\ref{GBD})) against the scale factor $a$ in Model I.} \label{G1}\end{minipage}\end{figure}

Hence, from the second field equation, we obtain that the BD parameter $\omega$ can be written as:
\begin{eqnarray}\label{51}
\omega_{\pm}(\phi)=(4q-2)(\alpha_{\pm}a)^{-2}+(2q-4)(\alpha_{\pm}a)^{-1}-2.
\end{eqnarray}
We have obtained the scalar field $\phi$ as a function of the scale factor $a$ and the BD parameter $\omega$ as a function of the scale factor $a$ and the deceleration parameter $q$. We shall now make a pictorial presentation of the scalar field $\phi$ against the time$t$ and the scale factor $a$, the BD parameter $\omega$ against the time $t$ and $G=1/\phi$ against the scale factor $a$. In the figures we shall consider $H_0t_0=0.95,~f_2=n/f_1^2,~f_3=f_0t_0^2$. The blue and green line indicate, respectively, $\phi_{+}$ and $\phi_{-}$ . The solid line corresponds to $n=0.75$ while the dashed line corresponds to $n=0.2$ in Figs. \ref{phi}, \ref{phia1} and \ref{omega}. In Fig. \ref{phi}, we have plotted $\phi_{\pm}$ against the time $t$. We observed an increasing pattern of the scalar field $\phi$ (blue line) for $\alpha_{+}$ and a decreasing pattern (green line) for $\alpha_{-}$. Similarly, in Fig. \ref{phi}, we have plotted $\phi_{\pm}$ against the scale factor $a$ and we observed an increasing pattern of the scalar field $\phi$ (blue line) for $\alpha_{+}$ and a decreasing pattern (green line) for $\alpha_{-}$.  In Fig. \ref{omega}, we can clearly see that the BD parameter $\omega(\phi)$ is increasing with the time $t$ and it has negative values; moreover $\omega<-3/2$, which corresponds to cosmic acceleration (a result in agreement with already found results \cite{sharif1,sahoo}). However, this is incompatible with solar system constraints which require $\omega\geq 40,000$ \cite{sharif1}. This is the generic problem
noted in the context of scalar tensor theories as pointed out in the work of Sharif $\&$ Waheed \cite{sharif1}.  Combining Eqs. (\ref{8}) and (\ref{29}), we can express the scale factor $a$ in terms of the deceleration parameter $q$ as follow
\begin{eqnarray}\label{55}
a\left(q\right)&=&\left(\frac{\sqrt{n}-n}{H_0t_0-n}\right)^n\exp (\sqrt{n}-H_0t_0) \times \nonumber \\
&&\left[\frac{1}{\sqrt{n}-n}\left\{\left(\frac{q+1}{n}\right)^{-1/2}-n\right\}\right]^n\exp\left[\left(\frac{q+1}{n}\right)^{-1/2}-\sqrt{n}\right].
\end{eqnarray}
The expression of $a$ given in Eq. (\ref{55}), when substituted in Eqs. (\ref{47}) and (\ref{48}), gives the BD parameter in terms of the deceleration parameter $q$. Due to the presence of $t_0$, i.e. the age of the universe, the term $\frac{4f_3}{f_1^2}$ may be the most dominating term in Eq. (\ref{46}). This may lead to positive and negative values of $\alpha$. However, if $\phi_0$ is sufficiently small then the term $f_3$ will be large enough to make $\alpha_{+}\approx-\alpha_{-}$. As it is well known, for theories with constant $\omega$, the possibility of variations of $G$ is very small. Consideration of arbitrary coupling function $\omega(\phi)$, however, opens
the possibility of variations of $G$. According to Greenstein \cite{green}, the expression of $G$ in BD cosmology is defined as:
 \begin{eqnarray}\label{GBD}
 G(t)=\frac{1}{\phi(t)}.
\end{eqnarray}
In Fig. \ref{G1}, we have plotted $\frac{\dot{G}}{G}$ for the Model I. The choice of parameters is the same as the last three figures. The blue and green lines correspond, respectively, to $\alpha_{+}$ and $\alpha_{-}$. Moreover, the solid and the dotted lines correspond, respectively, to $n=0.80$ and $n=0.6$. We observe that, for $\alpha_{-}$, we have $\frac{\dot{G}}{G}>0$ while, for $\alpha_{+}$, we have $\frac{\dot{G}}{G}<0$. We further note that $\left|\frac{\dot{G}}{G}\right|$ is increasing in both cases. The rate of increasing is sharper for $n=0.60$ than for  $n=0.80$.
\subsection{Model II}
In this subsection, we consider the following expression of $\phi\left(a\right)$:
\begin{eqnarray}\label{56}
\phi\left(a\right)=\phi_0\exp[\beta a].
\end{eqnarray}
Combining Eqs. (\ref{31}) and (\ref{32}) and using in Eq. (\ref{56}), we get:
\begin{eqnarray}\label{57}
(1+\beta)\frac{\ddot{a}}{a}+(\beta^2+4\beta+4)H^2=\frac{\rho}{\phi},
\end{eqnarray}
which leads to (for the present epoch):
\begin{eqnarray}\label{60}
\beta_{\pm}=\frac{q_0-4\pm\sqrt{q_0^2+16+\frac{4\rho_0t_0^2}{\phi_0(H_0t_0)^2}}}{2}=\frac{f_2-5\pm\sqrt{f_2^2-2f_2+4f_3/f_1^2+17}}{2}.
\end{eqnarray}
Using Eq. (\ref{56}) in Eq. (\ref{32}) and considering $\beta=\frac{1}{a}\ln \left(\frac{\phi}{\phi_0}\right)$, we get $\omega$ in terms of the deceleration parameter $q$ as follow:
\begin{eqnarray}\label{61}
\omega \left(\phi\right)=(4q-2)\left[\frac{1}{a}\ln \left(\frac{\phi}{\phi_0}\right)\right]^{-2}+2(q-1)\left[\frac{1}{a}\ln \left(\frac{\phi}{\phi_0}\right)\right]^{-1}-2.
\end{eqnarray}
Using Eqs. (\ref{56}) and (\ref{60}), we get:
\begin{eqnarray}\label{63}
\phi_{\pm}&=&\phi_0\exp\left[\frac{q_0-4\pm\sqrt{q_0^2+16+\frac{4\rho_0t_0^2}{\phi_0(H_0t_0)^2}}}{2}a\right] \nonumber \\
&=& \phi_0\exp\left[\frac{f_2-5\pm\sqrt{f_2^2-2f_2+4f_3/f_1^2+17}}{2}a\right].
\end{eqnarray}

\begin{figure}[ht] \begin{minipage}[b]{0.45\linewidth} \centering\includegraphics[width=\textwidth]{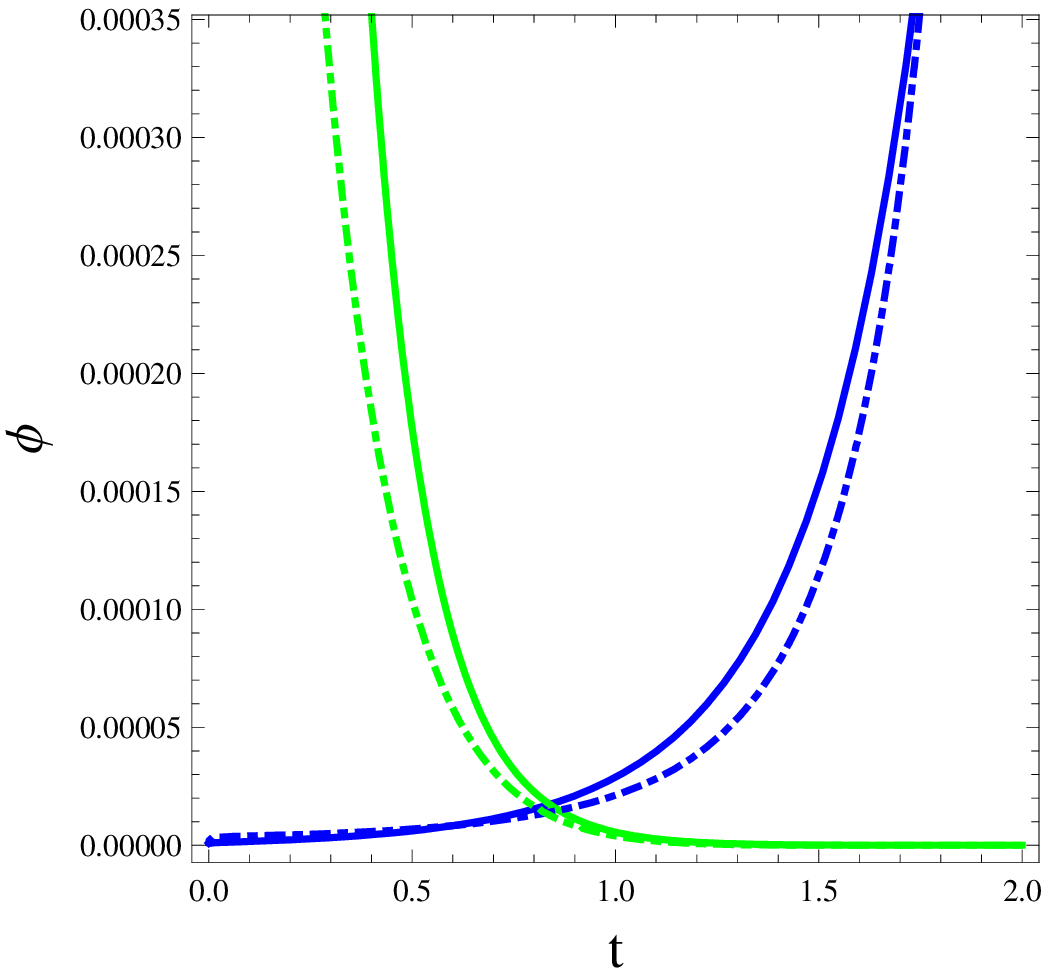} \caption{Plot of $\phi$ (see Eq. (52)) against the cosmic time $t$ in Model II.} \label{phi2} \end{minipage} \hspace{0.5cm} \begin{minipage}[b]{0.45\linewidth} \centering\includegraphics[width=\textwidth]{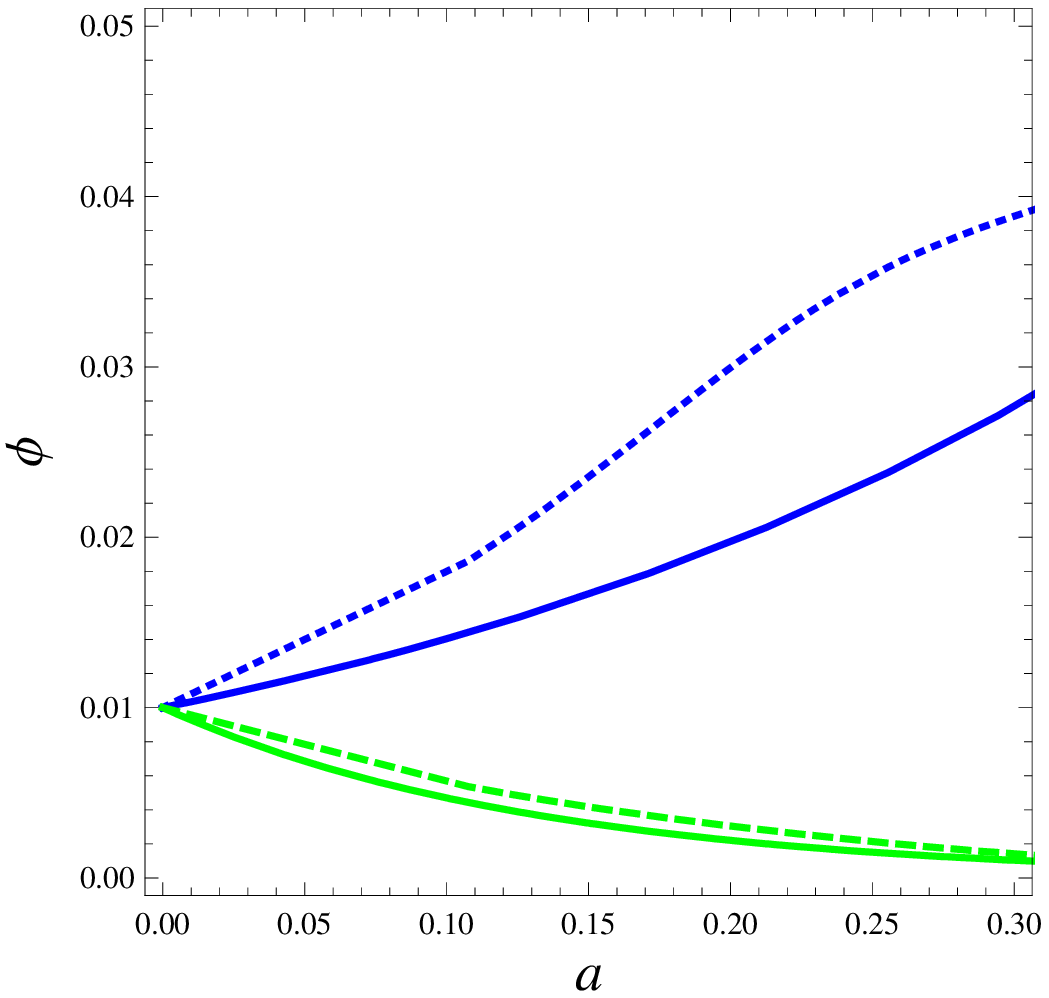} \caption{Plot of $\phi$ (see Eq. (52)) against the scale factor $a$ in Model II.} \label{phia2} \end{minipage} \hspace{0.5cm} \begin{minipage}[b]{0.45\linewidth} \centering\includegraphics[width=\textwidth]{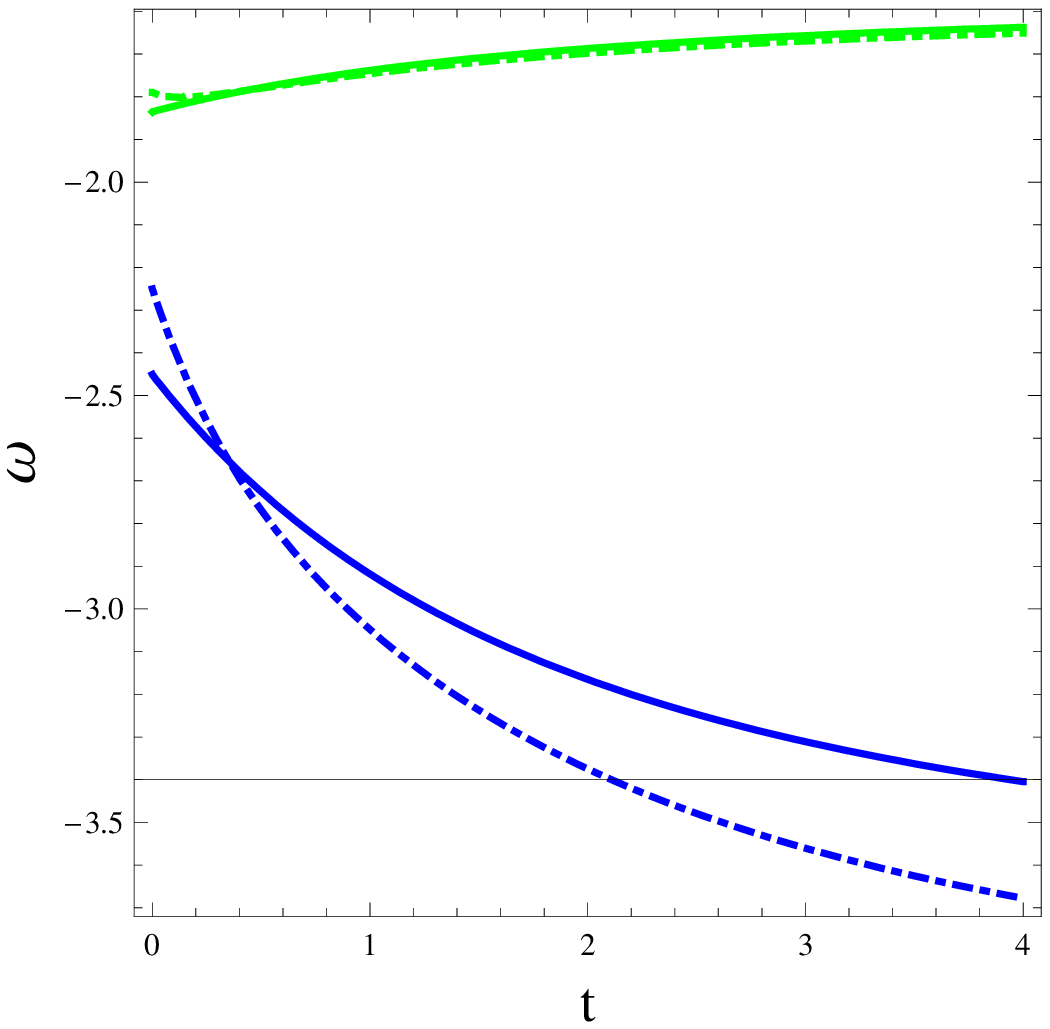}\caption{Plot of BD parameter $\omega$ (using Eqs. (51) and (52)) against the time $t$ in Model II.} \label{omega2} \end{minipage}\hspace{0.5cm}\begin{minipage}[b]{0.45\linewidth} \centering\includegraphics[width=\textwidth]{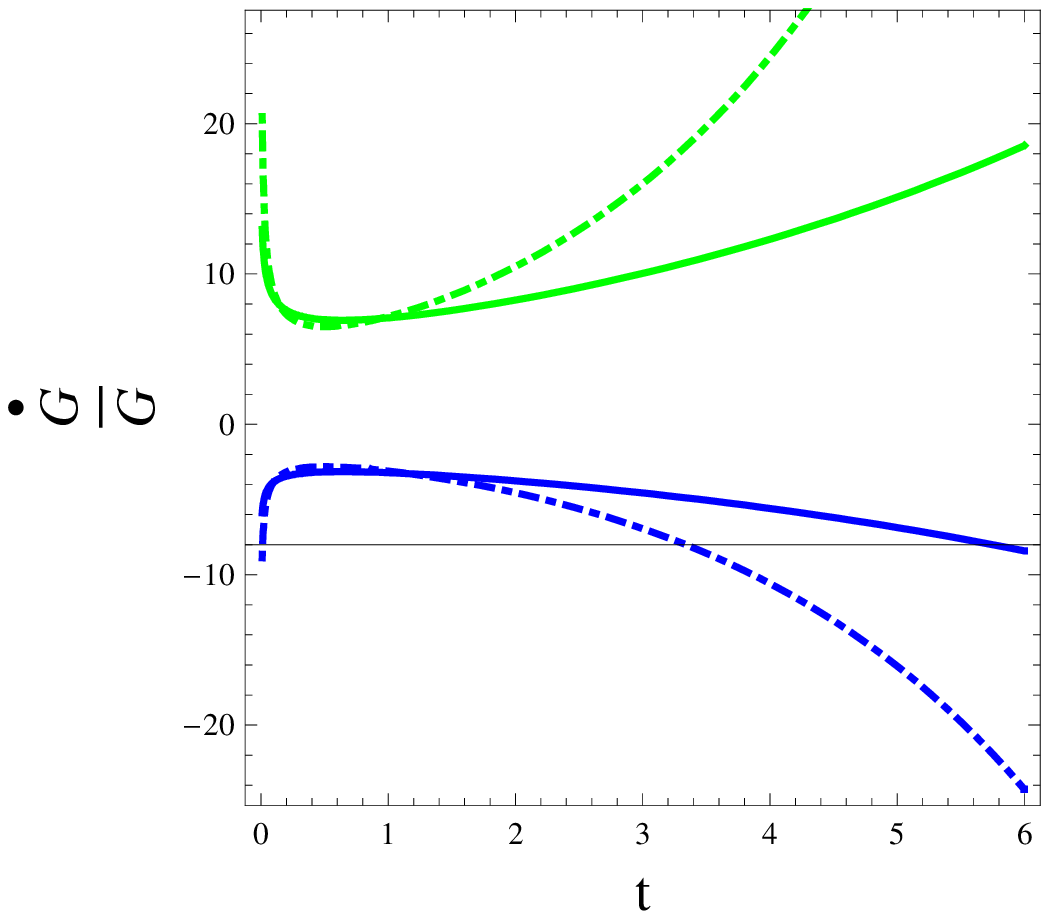}\caption{Plot of time derivative $\frac{\dot{G}}{G}$ (see Eq. (\ref{GBD})) against the time $t$ in Model II.} \label{G2}\end{minipage}\end{figure}

In Fig. (\ref{phi2}), we can observe that, for $n=0.75$, the scalar field $\phi$ is increasing with the time $t$ while, for $n=0.2$, $\phi$ is decreasing with the time $t$. Hence, this behavior is the same as Model I. However, in Fig. (\ref{omega2}), we observe a different pattern of the BD parameter from Model I. Although $\omega<-3/2$ holds for both values of $n$ like Model I, $\omega$ decreases for $n=0.75$ contrary to what happened in Model I. However, the difference in $\beta_{+}$ and $\beta_{-}$ does not have any impact on the patterns of the scalar field and the BD parameter. In Fig. (\ref{G2}), we have plotted $\frac{\dot{G}}{G}$ for the Model II. The choice of the parameters is the same as the last three Figures. The blue and green lines correspond, respectively, to $\alpha_{+}$ and $\alpha_{-}$ while the solid and the dotted lines correspond, respectively, to $n=0.80$ and $n=0.60$. We observe that, for $\beta_{-}$, we have $\frac{\dot{G}}{G}>0$ while, for $\beta_{+}$, we have $\frac{\dot{G}}{G}<0$. We further note that $\left|\frac{\dot{G}}{G}\right|$ is increasing in both cases. The rate of increasing is sharper for $n=0.60$ than for  $n=0.80$. In general, the behavior is almost similar to that of the Model I.
\\
\subsection{Model III}
If we combine Eqs. (\ref{31}) and (\ref{32}), we get:
\begin{eqnarray}\label{65}
\ddot{\phi}+5\left(\frac{\dot{a}}{a}\right)\dot{\phi}+2\left[2\left(\frac{\dot{a}}{a}\right)^2+\frac{\ddot{a}}{a}\right]\phi=\frac{\rho_0}{a^3}.
\end{eqnarray}
For the scale factor $a \left(t\right)=(t/t_0)^n\exp[b(t-t_0)]$, we get the solution for the scalar field $\phi$ from Eq. (\ref{65}) as follow:
\begin{eqnarray}\label{phi3}
\phi\left(t\right)=e^{-2bt}\left(C_1t^{-2n}-\frac{C_2t^{-3n}(bt)^n~\Gamma[1-n,bt]}{b}-\frac{e^{3bt_0}\rho_0(bt)^n\left(\frac{t}{t_0}\right)^{-3n}\Gamma[2-n,bt]}{b^2}\right),
\end{eqnarray}
where  $C_1$ and $C_2$ are two constants of integration and $\Gamma$ indicates the Gamma function. Using Eq. (\ref{phi3}) in  Eq. (\ref{31}), we can get:
\begin{eqnarray}\label{omega3}
\omega\left(t\right)=\frac{\psi_1}{\psi_2},
\end{eqnarray}
where:
\begin{eqnarray}
 \psi_1 &=& 2e^{bt}\left(-b^2C_1t^nT^{3n}\Gamma[1-n,bt]+e^{3bt_0}\rho_0t^{3n}(bt)^n\Gamma \left[2-n,bt \right] \right)\times \nonumber \\
  &&\left(b^2(-e^{3bt_0}\rho_0t^{2+3n}(-1+3n+3bt)-3C_2t(n+bt)T^{3n}+3C_1e^{bt}t^n(n+bt)^2T^{3n})-\right. \nonumber\\
 &&\left.3bC_2e^{bt}(bt)^n(n+bt)^2T^{3n}\Gamma \left[1-n,bt \right]-3e^{b(t+3t_0)}\rho_0t^{3n}(bt)^n(n+bt)^2\Gamma \left[2-n,bt\right]\right), \label{omega31}\\
 \psi_2 &=& \left[b^2\left(e^{3bt_0}\rho_0t^{2+3n}+C_2tT^{3n}-2C_1e^{bt}t^n(n+bt)T^{3n}\right)+2bC_2e^{bt}(bt)^n(n+bt)T^{3n}\times \right.\nonumber \\
 &&\left.\Gamma \left[1-n,bt\right]+2e^{b(t+3t_0)\rho_0t^{3n}(bt)^n(n+bt)\Gamma[2-n,bt]}\right]^2. \label{omega32}
 \end{eqnarray}

 \begin{figure}
\centering
\includegraphics[height=8.2cm]{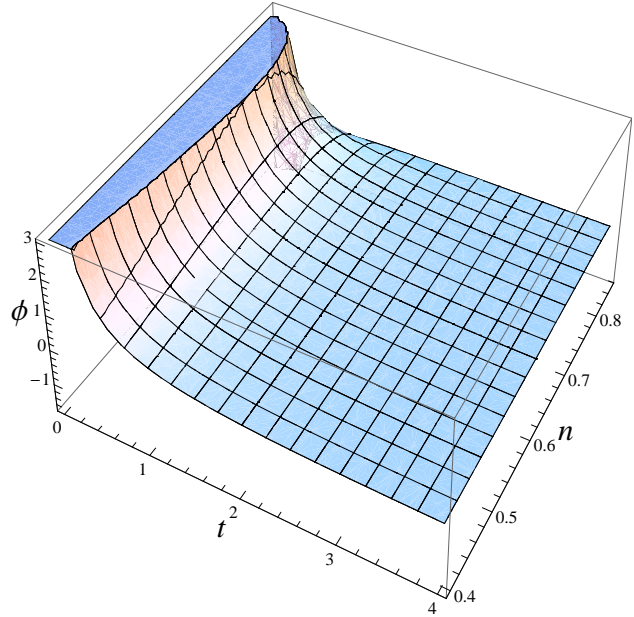}
\caption{Plot of $\phi$ against the time $t$ and $n$ from Eq. (\ref{phi3}).}
\label{phi3333}
\end{figure}
\begin{figure}
\centering
\includegraphics[height=8.2cm]{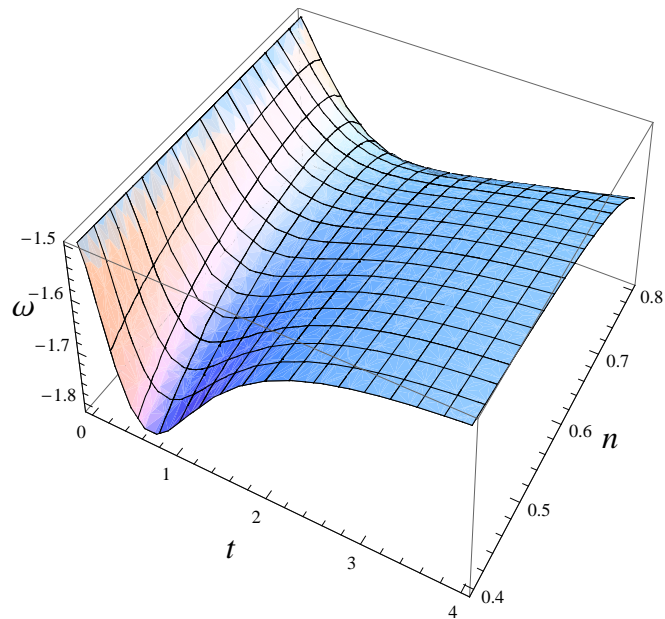}
\caption{Plot of $\omega$ against the time $t$ and $n$ from Eq. (\ref{omega3}).}
\label{omega3333}
\end{figure}
\begin{figure}
\centering
\includegraphics[height=8.2cm]{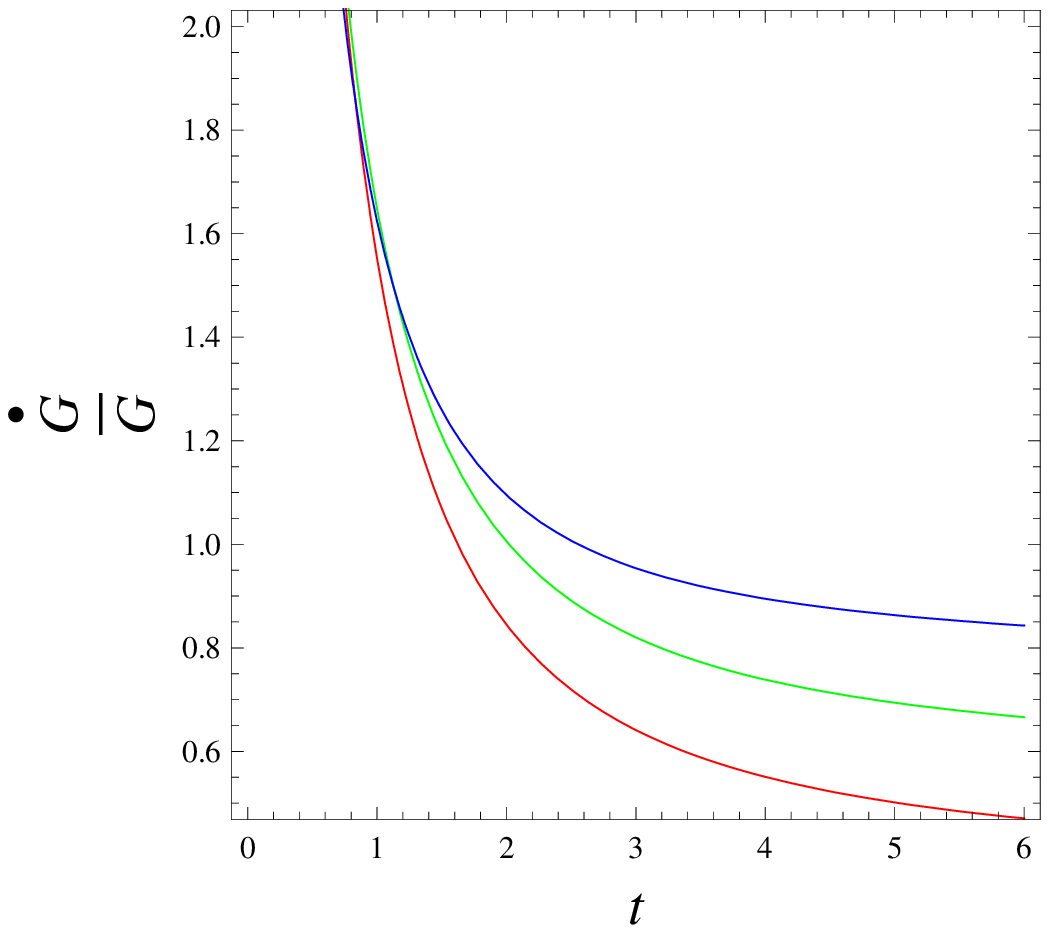}
\caption{Plot of $\frac{\dot{G}}{G}$ against the scale factor $a$ for $0<n<1$ from Eq. (\ref{phi3}).}
\label{G3333}
\end{figure}
In above equations, $T=\frac{t}{t_0}$ is used for the sake of convenience.
In Fig. \ref{phi3333}, we have plotted the scalar field based on Eq. (\ref{phi3}). We have taken $b$ from $b=\frac{\sqrt{n}-n}{t_1}$, where $t_1=\frac{(\sqrt{n}-n)t_0}{H_0t_0-n}$. Like the previous plots, we have taken $H_0t_0=0.95$. We have observed a decreasing pattern of the scalar field $\phi$ with the evolution of the universe. In Fig. \ref{omega3333}, we have plotted the BD parameter $\omega$ based on Eq. (\ref{omega3}). The parameters are chosen in the same way as before. Here, $\omega\leq-3/2$ holds for all of the values on $n$ under consideration. With increase in the value of $n$, $\omega$ is tending to $-3/2$. In Fig. \ref{G3333} we have plotted $\frac{\dot{G}}{G}$ against the time $t$ with for $n=0.80~\textrm{(red)},~0.70~\textrm{(green)}$ and $0.60~\textrm{(blue)}$. We observed that $\frac{\dot{G}}{G}$ is positive and decreasing with the time $t$. Since $\phi \left(t\right)$ is positive and $G=\frac{1}{\phi}$ we must have $\dot{G}>0$. According to \cite{green}, models with $\dot{G}>0$ at the present epoch produce no primordial $He^4$ and have ages significantly lower than the corresponding relativistic ages.

An important point to mention here is that the time variation
of $G$ does not directly affect the nuclear process of the early universe. But the expansion rate of the universe in this
type of theory do influence the primeval nucleosynthesis \cite{anjan}. It is already stated that the models with $\dot{G}>0$ at the present epoch produce no primordial $He^4$ and have ages significantly lower than the corresponding relativistic ages. If $H_0t_0>1$, then universe is always accelerating which seriously contradicts the nucleosynthesis scenario. One way to avoid such
problem is to consider $\omega$ as a function of the scalar field $\phi$. In
a recent work, Banerjee $\&$ Pavon \cite{NB1} have shown that, with
$\omega \left(\phi \right)$, one can have a decelerating radiation dominated era in
the early time and accelerated matter dominated era in the late
time. Also in the present work, we have followed the way of \cite{NB1} in a matter dominated universe.

In Table I we have presented $\left|\frac{\dot{G}}{G}\right|_{a=1}$ to examine viability of the models. It is well documented in the literature that $\left|\frac{\dot{G}}{G}\right|_{a=1}\leq4\times10^{-10}yr^{-1}$. For Models I and II with $\alpha_{+}$ and $\beta_{+}$ and Model III, the computed $\left|\frac{\dot{G}}{G}\right|_{a=1}$ lie well within the allowed range of variation of $G$. Hence, we discard Models I and II with $\alpha_{-}$ and $\beta_{-}$ as they produce range of variation of $G$ outside the range of its allowed limit for cosmic acceleration.

\begin{table}[ht]
\caption{Values of $\left|\frac{\dot{G}}{G}\right|_{a=1}$ for $n=0.80$. } % title of Table
\centering % used for centering table
\begin{tabular}{c c c c} % centered columns (4 columns)
\hline\hline %inserts double horizontal lines

Model & Case & $\left|\frac{\dot{G}}{G}\right|_{a=1}$ & Observation \\ [0.5ex] % inserts table
%heading
\hline % inserts single horizontal line
Model I & $\alpha_{+}$ & $2.51\times10^{-10}yr^{-1}$  & $<4\times10^{-10}yr^{-1}$ \\ % inserting body of the table
Model I & $\alpha_{-}$  & $7.53\times10^{-10}yr^{-1}$ & $>4\times10^{-10}yr^{-1}$ \\
Model II & $\beta_{+}$ & $3.19\times10^{-10}yr^{-1}$ & $<4\times10^{-10}yr^{-1}$ \\
Model II & $\beta_{-}$ & $7.11\times10^{-10}yr^{-1}$ & $>4\times10^{-10}yr^{-1}$ \\
Model III & -- & $1.54\times10^{-10}yr^{-1}$  & $<4\times10^{-10}yr^{-1}$ \\ [1ex] % [1ex] adds vertical space
\hline %inserts single line
\end{tabular}
\label{table:nonlin} % is used to refer this table in the text
\end{table}

\section{Concluding remarks}
The present study is motivated by the work of Ganguly $\&$ Banerjee \cite{NB}, who have shown that by expressing the dimensionless parameter $\omega$  in the Brans-Dicke theory as a function of the scalar field $\phi$ in a certain way, the process of expansion of the universe can be shown to make a transition from an initial phase of deceleration to a phase of acceleration, manifested through a signature flip of the deceleration parameter $q$ at some instant of cosmic time. In the first phase of the present study we have chosen the scale factor $a \left(t\right)$ in such a way that the deceleration parameter $q$, based on it, evolves into a negative value from a positive one as a function of time. Our choice is $a \left(t\right)=At^n\exp[bt]$. The functional dependence between $a \left(t\right)$ and $q \left(t\right)$ have been analyzed in details. Based on our chosen scale factor, the Hubble parameter $H$ has been determined as a function of time. In the second phase of this study, using Brans-Dicke theory, we have derived an expression of the parameter $\omega$ as a function of the scalar field $\phi$. The time dependence of $\phi$ and $\omega \left(\phi\right)$ has been determined on the basis of the chosen scale factor and the subsequent deceleration parameter of our model. In the second phase of the study we have considered three models. In the first model we have chosen scalar field $\phi$ as a function of $a$ as $\phi(a)=\phi_1\exp[\alpha a]$. Using the modified field equations for Brans-Dicke theory (Eqs. (\ref{31}) and (\ref{32})) with this choice of scalar field we have obtained a quadratic equation of $\alpha$ and for both of its roots we have observed the behaviors of $\phi$ against the scale factor $a$ and the time $t$. Moreover, we have studied the behavior of $\omega$ as a function of the time $t$. We have seen that $\omega \left(\phi \right)$ is increasing with time $t$, it has negative values and $\omega<-3/2$, which corresponds to cosmic acceleration and in agreement with already found results. Moreover, we have studied the behavior of $\frac{\dot{G}}{G}$, where $G(\phi)=\frac{1}{\phi}$.  We have observed that for one roof ($\alpha_{-}$) we have $\frac{\dot{G}}{G}>0$ and for another root ($\alpha_{+}$) we have $\frac{\dot{G}}{G}<0$. We further have noted that $\left|\frac{\dot{G}}{G}\right|$ is increasing in both cases. The rate of increasing is getting sharper with decrease in the value of $n$. In the next model we have considered the scalar field as $\phi \left(a \right)=\phi_0\exp[\beta a]$. Proceeding in the same manner as in the previous model we have obtained a quadratic equation of $\beta$ and for both of its roots we have observed the behaviors of $\phi$ against the scale factor $a$ and the time $t$. We have observed similar behavior as of the first model. Further we have observed that for the root $\beta_{-}$ we have $\frac{\dot{G}}{G}>0$ and for $\beta_{+}$ we have $\frac{\dot{G}}{G}<0$. Comparing with the observational limit of $\left|\frac{\dot{G}}{G}\right|$ at $a=1$, we have discarded $\alpha_{-}$ and $\beta_{-}$. The valid roots are found to be $\alpha_{+}=\frac{f_2-6+\sqrt{f_2^2-4f_2+4f_3/f_1^2+12}}{2}$ and $\beta_{+}==\frac{f_2-5\pm\sqrt{f_2^2-2f_2+4f_3/f_1^2+17}}{2}$. In the said models we assumed two different ansatz for the scalar field $\phi$. In the third model we did not make any assumption regarding the scalar field $\phi$. Using the the chosen scale factor in the field equations we have obtained the scalar field and Brans-Dicke parameter as functions of $t$ and based on them observed the behaviours of $\phi$, $\omega$ and $\frac{\dot{G}}{G}$. $\left|\frac{\dot{G}}{G}\right|_{a=1}$ is found to be within the allowed range.

A question may arise about the necessity of considering two ansatz for the scalar fields when analytical solution for $\phi$ exists as a function of $t$. In Models I and II we have tried to investigate the effects of particular choices of scalar field in Brans-Dicke theory with the above choice of scale factor. One clear finding is that for the choices of scalar field in Models I and II, the scalar field $\phi(t)$ is always an increasing function
of time in all epochs for the accepted roots $\alpha_{+}$ and $\beta_{+}$. However, the scalar field obtained without any assumption shows a monotonic decreasing behavior with time. For small scale factors $a$ i.e. at higher redshifts, $\omega \left(t\right)$ for all Models I and II are in close agreement with that of the Model III. Thus, at very early stages of the universe, choice of the scalar field does not have any significant impact on $\omega(t)$. However, in later stages $a>1.5$, Models I and II produced decreasing $\omega(t)$ and Model III produced increasing $\omega(t)$. Moreover, for Model III, $\omega(t)$ is not monotonic. Whereas, for the other models, it is strictly monotonic decreasing.

\begin{center}
\textbf{Acknowledgement}
\end{center}

Constructive suggestions from the reviewer is thankfully acknowledged by the author. The second author acknowledges financial support from the Department of Science and Technology, Govt. of India, under the Grant No. SR/FTP/PS-167/2011.

\end{document}